\documentclass[12pt]{article}
\usepackage{amsmath}
\usepackage{amsfonts}
\usepackage{amssymb}   
\usepackage{bm}             
\usepackage{graphicx}
\usepackage[margin=1in]{geometry}
\usepackage[doublespacing]{setspace}
\usepackage{subfigure}
\usepackage{multirow}
\usepackage[style=authoryear, natbib=true, url=false, isbn=false, doi=false, backend=bibtex]{biblatex}
\usepackage{color}
\bibliography{CR-MCMC-bibliography}
\renewbibmacro{in:}{}

\newcommand{\given}{\, | \,}

\begin{document}

\title{Efficient Markov Chain Monte Carlo Sampling for Hierarchical Hidden Markov Models}
\author{Daniel Turek$^{*}$, Perry de Valpine, and Christopher J. Paciorek}
\date{}
\maketitle
\thispagestyle{empty}

\vspace{0.2in}

\begin{center}
$^*$Corresponding author \\
University of California, Berkeley \\
493 Evans Hall, Berkeley, CA 94720, USA \\
dturek@berkeley.edu \\
\end{center}

\vspace{0.5in}

\begin{abstract}   
Traditional Markov chain Monte Carlo (MCMC) sampling of hidden Markov models (HMMs) involves latent states underlying an imperfect observation process, and generates posterior samples for top-level parameters concurrently with nuisance latent variables.  When potentially many HMMs are embedded within a hierarchical model, this can result in prohibitively long MCMC runtimes.  We study combinations of existing methods, which are shown to vastly improve computational efficiency for these hierarchical models while maintaining the modeling flexibility provided by embedded HMMs.  The methods include discrete filtering of the HMM likelihood to remove latent states, reduced data representations, and a novel procedure for dynamic block sampling of posterior dimensions.  The first two methods have been used in isolation in existing application-specific software, but are not generally available for incorporation in arbitrary model structures.  Using the NIMBLE package for R, we develop and test combined computational approaches using three examples from ecological capture-recapture, although our methods are generally applicable to any embedded discrete HMMs.  These combinations provide several orders of magnitude improvement in MCMC sampling efficiency, defined as the rate of generating effectively independent posterior samples.  In addition to being computationally significant for this class of hierarchical models, this result underscores the potential for vast improvements to MCMC sampling efficiency which can result from combinations of known algorithms.
\end{abstract}   

\vspace{0.5in}

\textbf{Keywords:} \\
\indent \indent Capture-recapture, Effective sample size, Hidden Markov model, Hierarchical model, MCMC, NIMBLE, Sampling efficiency

\thispagestyle{empty}
\newpage

\section{Introduction}

Hidden Markov models (HMMs) are widely applied for the analysis of time series data with incomplete or noisy observations together with stochastic system dynamics \citep{cappe2006inference, elliott2008hidden}.  HMMs are used in a diverse range of application domains, with recent attention in areas of speech recognition and natural language processing \citep{gales2008application}.  See \citet{macdonald1997hidden} for a broad review of HMM applications in disciplines such as as medicine, finance, sociology, and climatology.

For a single discrete HMM, likelihood calculation involves summing over the distribution of a sequence of unknown latent states.  This can be implemented either using standard direct filtering summations \citep[\emph{e.g.},][chapter 2]{elliott2008hidden} as part of either maximum likelihood or Bayesian analysis, or using Markov chain Monte Carlo \citep[MCMC;][]{Gilks2005, Brooks2011} for Bayesian analysis.  In the case of MCMC, the unknown state variables are included in MCMC sampling.  However, it is often the case that one or more HMMs are embedded in a larger hierarchical model, perhaps accounting for explanatory variables of state transition probabilities or shared variation among multiple time series.  In such cases practitioners may rely on MCMC to perform a Bayesian analysis, but they face a quandary of computational efficiency.  If they use standard MCMC software, they often have no choice to but to include the unknown latent state variables in MCMC sampling.  For large models this can contribute hundreds or thousands of dimensions which require MCMC sampling, to the point of rendering this approach computationally impractical.

In theory there are computational tradeoffs between using MCMC and direct filtering summation when embedding HMMs in a larger hierarchical model, but these tradeoffs have not been explored to date.  Here we do so, by considering combinations of several existing computational methods for fitting HMMs.  These methods include direct filtering to remove latent variables, using a reduced representation of observational data, and dynamic blocking of model parameters to achieve efficient MCMC sampling.  We demonstrate that for large models, a combination of these techniques can yield several orders of magnitude improvement in sampling efficiency.  This can make the analysis of such models practical, opening new possibilities for fitting complex hierarchical models.

As examples we draw upon capture-recapture and from ecological statistics \citep[for a broad review, see][]{lebreton2009modeling}.  In capture-recapture, each animal in a study generates a capture history over multiple observational periods.  These data can be modeled using discrete HMMs, where latent states may simply represent ``alive" or ``dead", or in the case of multistate capture-recapture, are more detailed such as including reproductive status or location.  We present a series of three examples of increasing complexity to study the tradeoffs in computational cost and MCMC mixing of several methodological approaches.  Our examples include a simple Cormack-Jolly-Seber capture-recapture model (``Dipper"), a simple multistate model (``Orchid"), and a larger multistate model with thousands of embedded HMMs (``Goose").

Some of the techniques we study are already supported in existing software, however only for specific applications or particular hierarchical structures.  The standalone program MARK \citep{white1999program} is perhaps the industry leader for applied capture-recapture.  MARK provides an application-specific MCMC algorithm for fitting multistate random effects capture-recapture models, which implements filtering over latent states to directly calculate model likelihoods.  MARK also supports a reduced representation of datasets with repeated observations -- known as an ``m-array" in capture-recapture -- however only for band-recovery analyses \citep{brownie1985statistical}.  More recently, M-SURGE \citep{choquet2004m} was developed specifically for multistate capture-recapture.  M-SURGE supports numerical integration to remove latent states, although this is used exclusively for maximum likelihood estimation, and never in combination with MCMC.  Furthermore, neither of these software programs expose these computational techniques for user control, nor are they applicable outside the domain of ecological capture-recapture.

We make use of the NIMBLE software for specifying hierarchical models and statistical algorithms \citep{nimble-software:2015} to generalize these computational approaches for embedded HMMs.  We consider particular combinations of techniques using the flexible and transparent algorithmic control provided by NIMBLE.  Although we draw upon capture-recapture for examples, our advances in efficient handling of HMMs can be embedded in any larger hierarchical model structure using NIMBLE.  However, we focus attention on the computational methodologies rather than implementation details.  For comparisons of interest we also include the widely used JAGS package \citep{Plummer2003} for MCMC.

\section{Computational Approaches to Discrete HMMs}

We begin with a general specification of discrete HMMs, and explain how multistate capture-recapture models may be framed in this context.  We then provide the model likelihood, and present a variety of approaches to computing it in the context of MCMC estimation.

\subsection{Discrete HMMs and Multistate Capture-Recapture}

Let $y_i = (y_{i1}, \ldots, y_{ik})$ represent the $i^\text{th}$ sequence of observations taken over sampling occasions $t = 1, \ldots, k$.  Each $y_{it} \in \mathcal{Y}$, where $\mathcal{Y}$ is the finite set of possible observations.  Similarly, let $x_i = (x_{i1}, \ldots, x_{ik})$ be the sequence of true underlying states at occasions $t = 1, \ldots, k$, with $x_{it} \in \mathcal{X}$ for finite set of states $\mathcal{X}$.  We will consider a total of $n$ observed sequences, hence the full data set is $y = (y_1, \ldots, y_n)$.  Finally, let $\theta$ be a vector of all model parameters, which may also include random effects.  Letting $i$ take all values in $1, \ldots, n$, the general hierarchical model is
\begin{equation} \label{eqn:HMM}
\begin{split}
\Theta & \sim p(\theta) \\
X_{i1} & \sim f_{i1}(x_{i1} \given \theta) \\
X_{it} \given X_{i,t-1} & \sim f_{it}(x_{it} \given \theta, x_{i, t-1}), \hspace{12pt}  t = 2, \ldots, k \\
Y_{it} \given X_{it} & \sim g_{it}(y_{it} \given \theta, x_{it}),          \hspace{26pt} t = 1, \ldots, k \\
\end{split}
\end{equation}

Here $p(\cdot)$ a prior distribution for parameter vector $\theta$, which may itself have one or more levels of stochastic interdependence. The distribution of each HMM initial state $x_{i1}$ is $f_{i1}(\cdot \given \theta)$.  Markov state transition probabilities are given by $f_{it}(\cdot \given \theta, x_{i, t-1})$ and observation probabilities by $g_{it}(\cdot \given \theta, x_{it})$.

Discrete HMMs have long been applied in the area of ecological capture-recapture \citep[\emph{e.g.},][]{Gimenez2007, king2012review, langrock2012flexible}. In this context, a set of $n$ distinct animals is monitored for $k$ sampling occasions.  Each $y_i$ represents the observation history of animal $i$, for $i = 1, \ldots, n$, which can be modeled using HMMs as in~(\ref{eqn:HMM}).  The set of possible observations $\mathcal{Y}$ may include a state to represent ``unobserved".  Since all $n$ animals are not typically observed on occasion $t=1$, each embedded HMM will ``begin" at the sampling period corresponding to the first genuine observation of that animal.

\subsection{Model Likelihood}

We now provide the model likelihood for the general HMM formulation in~(\ref{eqn:HMM}), which is used in the Bayesian estimation procedures described next.  We begin with the likelihood contribution from a single observation history,
\begin{equation} \label{eqn:Li}
\begin{split}
L(\theta \given y_i) & = \sum_{x_i \in \mathcal{X}^k} f_{i1}(x_{i1} \given \theta) \left(\prod_{t=2}^{k} f_{it}(x_{it} \given \theta, x_{i,t-1})\right) \left(\prod_{t=1}^{k} g_{it}(y_{it} \given \theta, x_{it})\right), \\
\end{split}
\end{equation}
where $\mathcal{X}^k$ denotes the standard $k$-fold Cartesian product of $\mathcal{X}$.  Using the likelihood components in~(\ref{eqn:Li}), the total model likelihood of $y$ is
\begin{equation*} 
L(\theta \given y) = \prod_{i = 1}^n L(\theta \given y_i).
\end{equation*}

\subsection{Computational Approaches}

We now describe several computational approaches to applying Bayesian estimation to embedded HMMs.  These strategies will form the basis for our comparisons, using examples from capture-recapture.

\subsubsection*{MCMC for latent states and parameters}

One approach to Bayesian estimation is to perform MCMC sampling of both the model parameters and latent states; that is, to sample from the full posterior distribution $p(\theta, x \given y)$.  Doing so makes use of Bayes law in the form:
\begin{equation*}
\begin{split}
p(\theta, x \given y) & \propto p(\theta) \; \prod_{i = 1}^n \; p(x_i \given \theta) \; p(y_i \given \theta, x_i) \\
\end{split}
\end{equation*}

Using this approach the dimension of the MCMC sampling problem can be very large, since there can be up to $nk$ latent state variables.  Although we expect the MCMC update of each individual variable will be fast, since the algorithmic complexity is limited to that of standard MCMC sampling algorithms (\emph{e.g.}, Metropolis-Hastings), there can be a large number of latent states.  In addition to the computational cost, this can result in slow MCMC mixing for latent states and parameters.

\subsubsection*{Filtering over latent states with MCMC for parameters}

An alternate approach makes use of direct filtering to calculate the likelihood contribution of each observation history.  This approach relies on the discrete HMM structure underlying each observed sequence $y_i$ in~(\ref{eqn:HMM}).  Doing so, we may perform MCMC sampling of the posterior distribution of $\theta$ only, rather than $(\theta, x)$ as in the latent state MCMC, and use filtering to calculate each $p(y_i \given \theta)$ as described in \citet{elliott2008hidden}.  The filtering MCMC approach makes use of Bayes law in the form:
\begin{equation} \label{eqn:filterBayes}
\begin{split}
p(\theta \given y) & \propto p(\theta) \; \prod_{i = 1}^n \; p(y_i \given \theta) \\
\end{split}
\end{equation}

For a general discrete HMM as specified in~(\ref{eqn:HMM}), the filtering likelihood calculation proceeds as follows.  Everything pertains to the $i^{th}$ observation history $y_i$ and we omit subscripts~$i$.  All probabilities are conditional on $\theta$, and we use $y_{1:t}$ to represent $y_1, \ldots, y_t$.  We begin by defining distributions for the latent state at each time step, and the conditional likelihood:
\begin{equation} \label{eqn:filterMCMC}
\begin{split}
P_t(x) & = \text{Pr}(X_t=x \given y_{1:t-1}) \\
& = \sum_{x_{t-1} \in \mathcal{X}} \, \text{Pr}(X_t=x \given X_{t-1}=x_{t-1}) \, \text{Pr}(X_{t-1}=x_{t-1} \given y_{1:t-1}) \\
\\
Q_t(x) &= \text{Pr}(X_t=x \given y_{1:t}) \\
& = \text{Pr}(X_t=x \given y_{1:t-1}) \, \text{Pr}(Y_t=y_t \given X_t=x) / \text{Pr}(Y_t=y_t \given y_{1:t-1}) \\
\\
L_t & = \text{Pr}(Y_t=y_t \given y_{1:t-1}) \\
& = \sum_{x_{t} \in \mathcal{X}} \, \text{Pr}(Y_t=y_t \given X_t=x_t) \, \text{Pr}(X_t=x_t \given y_{1:t-1}) \\
\end{split}
\end{equation}

Mapping the elements of $\mathcal{X}$ to the indices $\{1, 2, \ldots, |\mathcal{X}|\}$, a bijection, we express each $P_t$ and $Q_t$ as column vectors of length $|\mathcal{X}|$.  Define $|\mathcal{X}| \times |\mathcal{X}|$ state transition matrices $T_t$ as having $(i,j)$ element $\text{Pr}(X_t=i \given X_{t-1}=j)$.  Similarly, define $|\mathcal{Y}| \times |\mathcal{X}|$ observation matrices $Z_t$ with $(i,j)$ element $\text{Pr}(Y_t=i \given X_{t}=j)$.  The elements of each $T_t$ and $Z_t$ are defined by $f_t$ and $g_t$, respectively, from~(\ref{eqn:HMM}). We rewrite~(\ref{eqn:filterMCMC}) in matrix form as
\begin{equation} \label{eqn:filterMCMCmatrix}
\begin{split}
P_t & = T_t \, Q_{t-1}, \hspace{56pt} t \geq 2 \\
Q_t & = Z_t(y_t)^{\prime} * P_t \, / \, L_t, \hspace{15pt} t \geq 1 \\
L_t & = Z_t(y_t) \, P_t, \hspace{49pt} t \geq 1, \\
\end{split}
\end{equation}
where $A(i)$ is the $i^{th}$ row of $A$, $A^\prime$ denotes matrix transposition, and $*$ represents element-wise multiplication.  The initial latent state distribution $P_1$ is specified by $f_{1}$ from the model specification~(\ref{eqn:HMM}), and all other $P_t$, $Q_t$, and $L_t$ terms are iteratively calculated using~(\ref{eqn:filterMCMCmatrix}).  The desired likelihood is calculated as $L(\theta \given y) = L_{1} L_{2} \cdots L_{k}$.  In related works \citep[\emph{e.g.},][]{kery_bayesian_2012corrected} $T_t$ and $Z_t$ may be transposed, resulting only in notational changes.

A simplification of this filtering algorithm is possible for the case of single-state capture-recapture with one absorbing state.  Once an animal is deceased, it is guaranteed to remain in that state thereafter, where $\mathcal{X} = \{\text{``alive"}, \text{``dead"}\}$ and $\mathcal{Y} = \{\text{``seen"}, \text{``not seen"}\}$.  In this context we can express the likelihood of a capture history in terms of survival probabilities $\phi_t = \text{Pr}(X_t=\text{``alive"} \given X_{t-1}=\text{``alive"})$ and detection probabilities $p_t = \text{Pr}(Y_t=\text{``seen"} \given X_t=\text{``alive"})$ as
\begin{equation} \label{eqn:filterMCMCsimplified}
L(\theta \given y) = \left( \prod_{t=1}^{t_{\text{final}}-1} \phi_t \right) \left( \prod_{t=2}^{t_{\text{final}}} p_t^{y_t}(1-p_t)^{1-y_t} \right) \chi_{t_{\text{final}}},
\end{equation}
where we numerically assign $y_t=\text{``seen"}$ as $y_t=1$ and $y_t=\text{``not seen"}$ as $y_t=0$, $t_{\text{final}}$ is the time index of the final observed sighting (\emph{i.e.}, $t_{\text{final}} = \max \{t \given y_t = 1\}$), $\chi_k = 1$, and $\chi_{t} = 1 - \phi_t + \phi_t (1-p_t) \chi_{t+1}$ for $t < k$ \citep{Lebreton1992}.  Use of this simplified calculation for single-state capture-recapture will dramatically speed up likelihood evaluations relative to~(\ref{eqn:filterMCMCmatrix}), since the likelihood is expressed in closed form.

These filtering algorithms numerically integrate over sequences of latent states to directly calculate model likelihoods, removing the need to perform MCMC sampling of these latent variables.  However, the MCMC sampling step for each component of $\theta$ now requires application of a filtering algorithm for each observed history $y_i$.  Thus, this approach reduces the dimensionality of the MCMC sampling problem, but at the cost of increased computational complexity of each MCMC iteration.

\subsubsection*{Filtering MCMC with a reduced representation of the dataset}

A further specialized approach arises when there are repeated instances of identical observation histories in the full observed dataset $y$.  That is, multiple distinct individuals exhibited identical observation histories over the $k$ observational periods.  Let $n^*$ be the number of unique observation histories in the original dataset $y$.  We define a reduced representation $(y^*, m^*)$, where $y^*$ contains the $n^*$ unique histories appearing in $y$.  An accompanying vector of multiplicities $m^*$ indicates how many times each unique history appears in the original dataset, where history $y^*_i$ occurs in $y$ a total of $m^*_i$ times, for $i = 1, \ldots, n^*$.  

Using this reduced representation, we can express~(\ref{eqn:filterBayes}) such that the likelihood of each unique observation history is calculated only once.  This computational approach makes use of Bayes law in the form:
\begin{equation} \label{eqn:reducedFilterMCMC}
\begin{split}
p(\theta \given y) = p(\theta) \; \prod_{i = 1}^{n^*} \; p(y^*_i \given \theta)^{m^*_i} \\
\end{split}
\end{equation}
Computing according to~(\ref{eqn:reducedFilterMCMC}) requires only $n^*$ applications of the filtering likelihood calculation, rather than $n$ applications when using the filtering MCMC approach on the full dataset.  We expect to this provide an approximate factor of $n/n^*$ improvement in computational efficiency relative to the filtering MCMC on the original dataset.

\subsubsection*{Filtering MCMC with block sampling}

As a final approach, we consider joint (a.k.a. block) MCMC sampling of model parameters \citep{Roberts1997}.  In the case of correlated posteriors, it is well known that block sampling of highly-correlated parameter dimensions can result in improved MCMC mixing \citep[\emph{e.g.},][]{Liu1994}.  The general problem of determining posterior dimensions for block sampling is difficult, as a practitioner cannot reliably guess what blocking arrangement will result in efficient MCMC sampling.  Further, existing literature on the efficiency of block sampling generally only considers the mixing properties of univariate versus block sampling, and fails to consider computational demands \citep[among others]{Mengersen1996, Roberts1996a, Roberts1997a}.

We make use of NIMBLE's automated procedure for determining an efficient problem-specific block sampling MCMC algorithm, which exemplifies how the flexibility and programmability of NIMBLE facilitates a higher level of algorithmic control than other statistical software packages.  This procedure dynamically determines a partition of the model parameters which results in efficient MCMC sampling.  MCMC efficiency is defined as the number of effectively independent posterior samples generated per second of algorithm runtime, which balances improvements in MCMC mixing with computational requirements.  This automated blocking procedure is described in detail in \citet{turek2015automated}.

The use of a block sampling strategy can be combined with filtering over latent states.  Under this approach we use the filtering algorithms already described to integrate out the latent states, and require MCMC sampling for the model parameters.  We use a dynamically determined block sampling strategy for the MCMC sampling of these parameters.

\section{Capture-Recapture Example Models}

We use three capture-recapture examples representing different levels of complexity to asses performance of the various computational approaches to MCMC estimation.  The first is the well-studied European Dipper dataset, demonstrating single-state capture-recapture.  The second is a multistate capture-recapture dataset of observations of a flowering orchid.  This is considered multistate data since the orchids may be observed in multiple distinct states, in addition to the possibility of ``not seen".  The third and largest dataset is also a multistate example, representing observations of Canadian Geese at various locations.

\subsection{Dipper Model}

The European Dipper (\emph{Cinclus cinclus}) dataset has been analyzed extensively in the literature \citep[][among numerous others]{marzolin1988polygynie, Lebreton1992, Gimenez2007, royle_modeling_2008, amstrup_handbook_2010}, and may be considered a canonical example of capture-recapture.  For simplicity, we do not make use of a covariate reflecting gender or the distinction of flood years as in \citet{Lebreton1992}.

The dataset consists of $n=294$ sighting histories collected over $k=7$ annual sighting occasions.  The set of latent states is $\mathcal{X} = \{\text{``alive"}, \text{``dead"}\}$ and the set of observable states is $\mathcal{Y} = \{\text{``seen"}, \text{``not seen"}\}$.  For computation, we use the numerical assignments $x=1$ for ``alive", $x=0$ for ``dead", $y=1$ for ``seen", and $y=0$ for ``not seen".

The model is parameterized by annual probability of survival, $\phi$, and probability of detection, $p$, which are assumed to be constant among all sampling occasions and individuals.  This reflects the most basic Cormack-Jolly-Seber model structure \citep{jolly1965explicit, seber1965note}, typically denoted as $\phi(.)~ p(.)$ to imply constant probabilities of survival and detection \citep[\emph{e.g.},][]{nichols1983estimation}.  The hierarchical model specification is given below, which is a realization of the general structure provided in~(\ref{eqn:HMM}), where $i$ assumes all values in $1, \ldots, n$.

\begin{equation*} 
\begin{split}
\phi & \sim \text{Uniform}(0, 1) \\
p & \sim \text{Uniform}(0, 1) \\
X_{i1} = Y_{i1} & = 1  \\
X_{it} \given X_{i,t-1} & \sim \text{Bernoulli}(\phi \; x_{i,t-1}) \hspace{20pt} t = 2, \ldots, k \\
Y_{it} \given X_{it} & \sim \text{Bernoulli}(p \; x_{it}) \hspace{34pt} t = 2, \ldots, k
\end{split}
\end{equation*}

\subsection{Orchid Model}

Our second example models sighting histories of the showy lady's slipper (\emph{Cypripedium reginae}), a flowering variety of orchid which is native to north America.  Here, the concept of ``capture" has been generalized to observational sightings.  One cannot observe these orchids with certainty due to a dormant state, in which the orchid is alive but not observable.

The Orchid model data consist of observational sighting histories of $n=250$ unique flowers, collected over $k=11$ annual observational periods.  There are four latent states, $\mathcal{X} = \{ \text{``vegetative"}, \text{``flowering"}, \text{``dormant"}, \text{``dead"} \}$, but only three distinct observable states, $\mathcal{Y} = \{ \text{``seen vegetative"}, \text{``seen flowering"}, \text{``not seen"} \}$ as we cannot distinguish between dormant and deceased flowers.  The presence of multiple distinct observable states (in addition to ``not seen") classifies this as multistate capture-recapture.  The full dataset is available in the supplementary material of  \citet{kery_bayesian_2012corrected}.

Following \citet{kery2004demographicCorrected} we include time-dependent survival probabilities $\phi_t$, and state transition probabilities $\psi_{rs}$ between the three living states.  We use an uninformative Dirichlet prior distribution for each set $\{ \psi_{1s}, \psi_{2s}, \psi_{3s} \}$, implemented using elemental $\text{Gamma}(1, 1)$ hyperpriors as in \citet{royle2008hierarchical}.  As flowers in the dormant state are never observed and there is no mis-identification of flowers in the vegetative or flowering states, the observation matrix $Z$ is deterministic.  In the model specification below, latent states $x_{it}$ are represented as binary column vectors, and $i$ assumes all values in $1, \ldots, n$.

\begin{equation*} 
\begin{split}
\phi_t & \sim \text{Uniform}(0, 1) \hspace{100pt} t = 2, \ldots, 11 \\
\{ \psi_{1s}, \psi_{2s}, \psi_{3s} \} & \sim \text{Dirichlet}(\alpha = \{1,1,1\} ) \hspace{52pt} s = 1, 2, 3 \\
X_{i1} & = y_{i1}  \\
X_{it} \given X_{i,t-1} & \sim \text{Categorical}(p = T_{t} \; x_{i,t-1}) \hspace{39pt} t = 2, \ldots, k \\
Y_{it} \given X_{it} & \sim \text{Categorical}(p = Z \; x_{it}) \hspace{54pt} t = 1, \ldots, k
\end{split}
\end{equation*}
which makes use of state transition matrices
\begin{equation*}
T_t = \left[
\begin{array}{cccc}
\phi_t \psi_{11} & \phi_t \psi_{12} & \phi_t \psi_{13} & 0 \\
\phi_t \psi_{21} & \phi_t \psi_{22} & \phi_t \psi_{23} & 0 \\
\phi_t \psi_{31} & \phi_t \psi_{32} & \phi_t \psi_{33} & 0 \\
1-\phi_t & 1-\phi_t & 1-\phi_t & 1
\end{array}
\right]
\end{equation*}
and constant observation matrix
\begin{equation*}
Z = \left[
\begin{array}{cccc}
1 & 0 & 0 & 0 \\
0 & 1 & 0 & 0 \\
0 & 0 & 1 & 1 \\
\end{array}
\right].
\end{equation*}

\subsection{Goose Model}

The multistate Goose model tracks $n=11,200$ Canadian Geese (\emph{Branta canadensis}) between three distinct locations over $k=4$ years.  Latent states $\mathcal{X} = \{ \text{``site A"}, \text{``site B"}, \text{``site C"}, \text{``dead"} \}$, with observable states $\mathcal{Y} = \{ \text{``seen at A"}, \text{``seen at B"}, \text{``seen at C"}, \text{``not seen"} \}$.  There exists a large number of identical sighting histories among the 11,200 geese, allowing a reduced representation using only the $n^*=153$ unique sighting histories.  The complete dataset can be found in \citet{amstrup_handbook_2010}.

Following \citet{amstrup_handbook_2010}, we include site-dependent survival probabilities, and both time- and site-dependent geographic transition probabilities and probabilities of detection.  We use uninformative priors for all parameters, including Dirichlet priors for each set of geographic transition probabilities.  Subsequent works \citep[\emph{e.g.},][]{mccrea2011multistate} have shown improved fits using more elaborate models for these data, but our purpose is to compare computational efficiency.  We desire high efficiency regardless of model fit, so the particular choice of model is tangential to our main points.  $i$ assumes all values in $1, \ldots, n$ in the hierarchical specification below.
\begin{equation*} 
\begin{split}
\phi_r & \sim \text{Uniform}(0, 1) \hspace{100pt} r = 1, 2, 3 \\
\{ \psi_{1st}, \psi_{2st}, \psi_{3st} \} & \sim \text{Dirichlet}(\alpha = \{1,1,1\} ) \hspace{52pt} s = 1, 2, 3, \hspace{10pt} t=2,3,4 \\
p_{rt} & \sim \text{Uniform}(0, 1) \hspace{101pt} r = 1, 2, 3, \hspace{10pt} t = 1, 2, 3, 4 \\
X_{i1} & = y_{i1} \\
X_{it} \given X_{i,t-1} & \sim \text{Categorical}(p = T_{t} \; x_{i,t-1}) \hspace{39pt} t = 2, \ldots, k \\
Y_{it} \given X_{it} & \sim \text{Categorical}(p = Z_t \; x_{it}) \hspace{51pt} t = 1, \ldots, k
\end{split}
\end{equation*}
which makes use of state transition matrices
\begin{equation*}
T_t = \left[
\begin{array}{cccc}
\phi_1 \psi_{11t} & \phi_2 \psi_{12t} & \phi_3 \psi_{13t} & 0 \\
\phi_1 \psi_{21t} & \phi_2 \psi_{22t} & \phi_3 \psi_{23t} & 0 \\
\phi_1 \psi_{31t} & \phi_2 \psi_{32t} & \phi_3 \psi_{33t} & 0 \\
1-\phi_1 & 1-\phi_2 & 1-\phi_3 & 1
\end{array}
\right]
\end{equation*}
and observation matrices
\begin{equation*}
Z_t = \left[
\begin{array}{cccc}
p_{1t}        & 0              & 0              & 0 \\
0               & p_{2t}       & 0              & 0 \\
0               & 0              & p_{3t}       & 0 \\
1-p_{1t}     & 1-p_{2t}   & 1-p_{3t}   & 1 \\
\end{array}
\right].
\end{equation*}

\section{Performance Results}

We now present the performance of various computational strategies for MCMC estimation applied to the three example capture-recapture models.  We do not present posterior results, but instead only the algorithmic efficiencies of each computational approach to generating these.  For each, the posterior results of top-level parameters closely agree with existing published analyses of the same datasets and models \citep{Lebreton1992, kery_bayesian_2012corrected, amstrup_handbook_2010}, which provides validation of our computational methodologies.

We include results for the following computational strategies MCMC estimation: latent state MCMC (``Latent State") where model parameters and latent states undergo MCMC sampling, filtering MCMC (``Filtering") in which we filter over latent states and only top-level parameters undergo MCMC sampling, and a combination of filtering and blocking (``Filtering \& Blocking") in which a customized blocking strategy is used for MCMC sampling of top-level parameters.  When appropriate, we also use a reduced representation (``RR") of the dataset.

We use the NIMBLE package for R to generate and execute MCMC algorithms, as the algorithmic flexibility it provides facilitates these computational approaches.  The use of user-defined distribution functions in NIMBLE allows us to incorporate the filtering algorithms (\ref{eqn:filterMCMCmatrix}) and (\ref{eqn:filterMCMCsimplified}) directly into a hierarchical model specification.  The generic discrete HMM filtering procedure described in (\ref{eqn:filterMCMCmatrix}) is used for filtering, or when permitted by the model structure we instead use the closed form likelihood calculation given in (\ref{eqn:filterMCMCsimplified}).  NIMBLE also provides the automated parameter blocking procedure \citep{turek2015automated} we use to generate problem-specific parameter blocking strategies for MCMC sampling.

We define the efficiency of an MCMC algorithm in terms of the number of effectively independent posterior samples produced per second of algorithm runtime.  This metric is denoted as effective samples per second (ESPS), and we will present both the minimum and mean ESPS among all model parameters.  This metric balances the tradeoff between computationally fast algorithms which generate highly autocorrelated chains of posterior samples, versus algorithms which are more computationally demanding but result in lower posterior autocorrelation, which provides stronger inferential power.

All algorithm runtimes represent the time required to generate 100,000 posterior samples.  When possible, we also provide comparisons with MCMC algorithms from the JAGS software package for R.  All calculations are produced using single-threaded execution on an Intel Xeon E5-2609 processor (2.40 GHz), running under the Ubuntu Linux operating system.

\subsection{Dipper Model}

\begin{figure}[h]
\centerline{\includegraphics[scale=1.0]{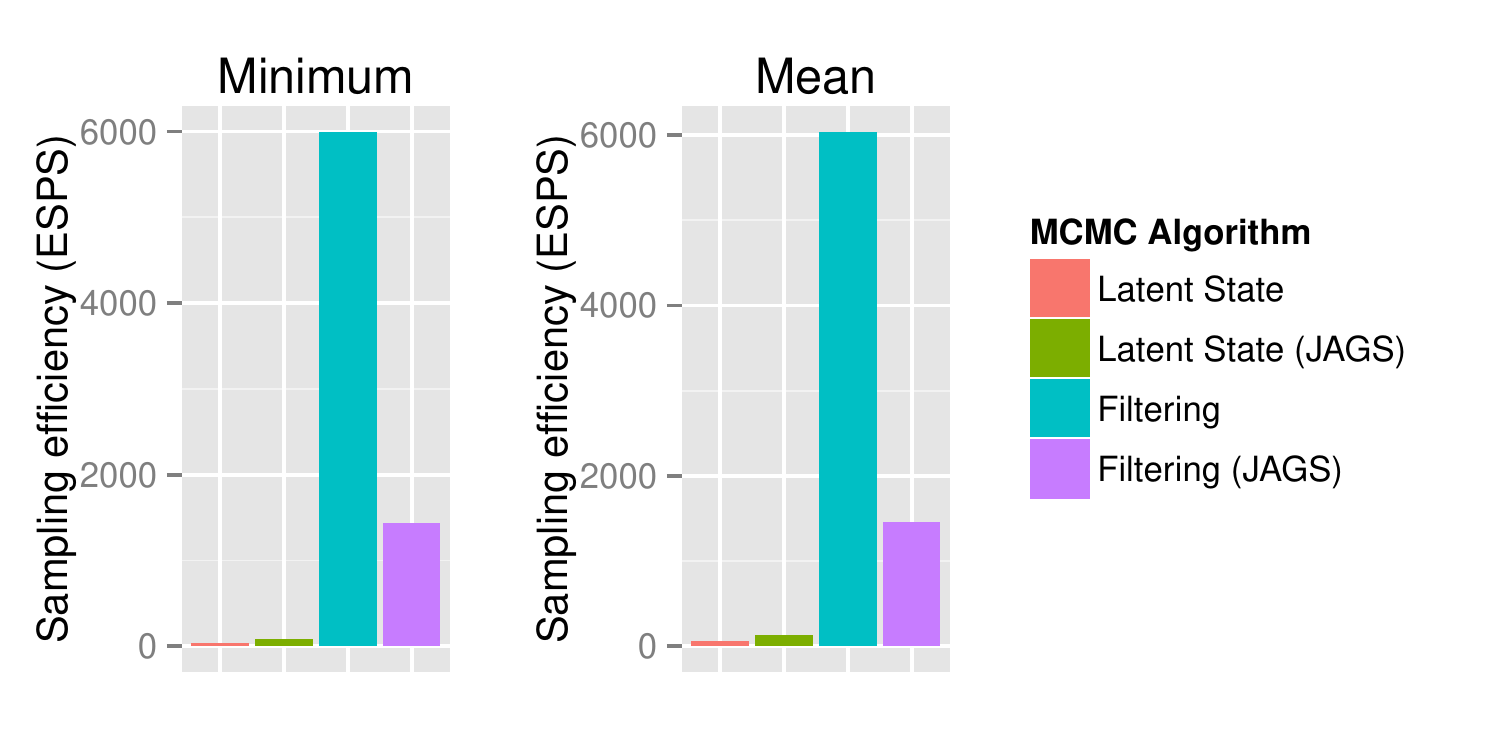}}
\caption{Minimum and mean parameter sampling efficiencies for the Dipper model.}
\label{fig:plotDipper}
\end{figure}

For the Dipper model, use of the filtering MCMC compared to MCMC sampling of all discrete latent states yielded a 60-fold improvement in sampling efficiency in NIMBLE and a 15-fold improvement in JAGS (Figure~\ref{fig:plotDipper}).  The sampling efficiencies of both top-level parameters are quite similar under each algorithm (although vary greatly between algorithms), hence the mean and the minimum summary statistics shown in Figure~\ref{fig:plotDipper} are similar as well.


The latent state MCMC requires MCMC sampling of 848 latent variables, in addition to the two top-level model parameters of interest.  The performance of JAGS is slightly better, although both result in sampling efficiencies of roughly 100 ESPS for both parameters.  NIMBLE and JAGS each require approximately four minutes to generate 100,000 samples.  The filtering MCMC is implemented in NIMBLE according to~(\ref{eqn:filterMCMCsimplified}), where only the two top-level parameters undergo MCMC sampling and runtime is reduced to 5 seconds.  The mixing also improves relative to the latent state MCMC, yielding a sampling efficiency of roughly 6,000 ESPS for both parameters, a 60-fold improvement.

For the Dipper model alone, we can also implement the filtering MCMC in JAGS.  This is possible because (\ref{eqn:filterMCMCsimplified}) provides a closed form expression for the likelihood of each sighting history.  This allows use of the ``zeros-trick" \citep[][p. 204-206]{Lunn2012} where a general log-likelihood expression is incorporated into a model through the mean parameter of a Poisson distribution, using an artificial zero-valued observation.  Using this technique reduces JAGS runtime to 30 seconds and increases sampling efficiency of both parameters to approximately 1,500 ESPS, a 15-fold improvement relative to the latent state MCMC.  Although the underlying calculations are similar to those of NIMBLE's filtering MCMC, this approach requires the additional overhead of artificial model variables and observations.

\subsection{Orchid Model}

\begin{figure}[h]
\centerline{\includegraphics[scale=1.0]{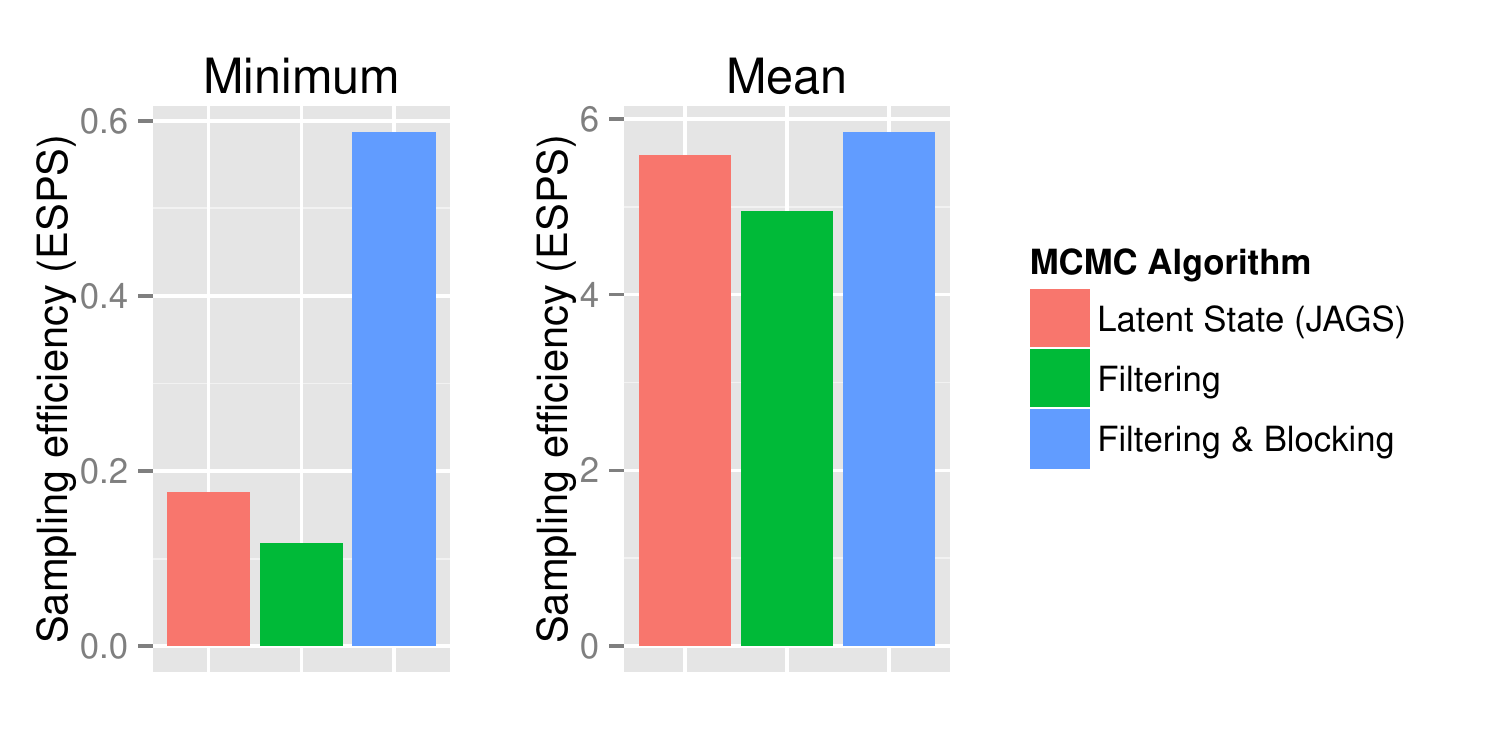}}
\caption{Minimum and mean parameter sampling efficiencies for the Orchid model.}
\label{fig:plotOrchid}
\end{figure}

For the multistate Orchid model, a combination of filtering over latent states and dynamic block sampling of parameters yielded a 3-fold improvement in sampling efficiency of the slowest mixing parameter, relative to the latent state MCMC (Figure~\ref{fig:plotOrchid}).


The latent state MCMC samples 2,157 latent variables in addition to 19 top-level parameters, which required 42 minutes to generate 100,000 samples.  Efficiency results for the latent state MCMC are quite similar to the filtering MCMC, which required 36 minutes but with slightly inferior mixing.  Both of these algorithms struggle to achieve good mixing among the nine state transition probabilities.  We might expect triplets of these parameters to be highly correlated due to the Dirichlet prior imposing a sum-to-one constraint, and indeed, examining the posterior correlations we find several instances of absolute pairwise posterior correlation greater than 0.9.  Under the latent state and filtering MCMC algorithms, several state transition probabilities have sampling efficiencies between 0.1 and 0.3 ESPS, which dictates the minimum efficiencies shown in Figure~\ref{fig:plotOrchid}.

For the 19 parameters undergoing MCMC sampling, NIMBLE's automated parameter blocking procedure converges on two blocks each consisting of two state transition probabilities, and univariate sampling for the other 15 parameters.  We observe that these pairs of transition probabilities have absolute posterior correlations of 0.98 and 0.97, the highest among all 19 parameters.  Joint sampling according to this blocking scheme in combination with filtering over latent states results in a minimum sampling efficiency of 0.6 ESPS, representing a 3-fold improvement over the latent state MCMC.




\subsection{Goose Model}

\begin{figure}[h]
\centerline{\includegraphics[scale=1.0]{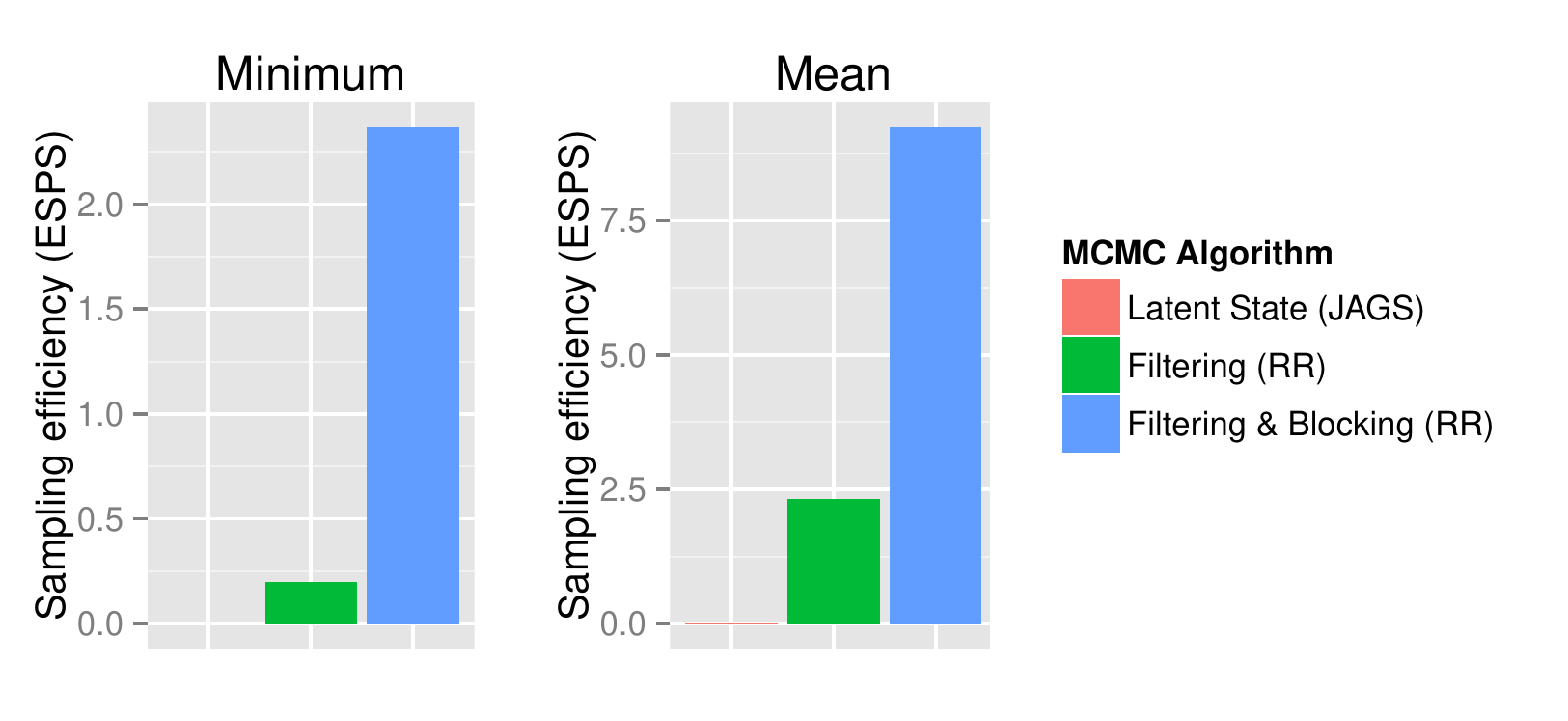}}
\caption{Minimum and mean parameter sampling efficiencies for the Goose model.}
\label{fig:plotGoose}
\end{figure}

As the Goose model includes a large number of repeated sighting histories among the 11,200 geese, this model benefits from a reduced representation of the data using the 153 unique sighting histories.  Applying the filtering MCMC to a reduced data representation produced a 70-fold improvement in sampling efficiency of the slowest mixing parameter, compared to the latent state MCMC (Figure~\ref{fig:plotGoose}).  An additional order of magnitude improvement was gained by applying dynamic blocking of model parameters.


The latent state MCMC requires sampling of 14,437 latent variables in addition to 21 top-level parameters.  We cannot use a reduced data representation under the latent state approach, since for correct inference each of the 11,200 sighting histories must have a corresponding sequence of latent state variables.  The latent state MCMC required approximately 24 hours to generate 100,000 samples, yielding a minimum sampling efficiency of 0.0027 ESPS and a mean of 0.028 ESPS.  This approach can be deemed impractical, as this translates to generating ten effective samples (for the slowest mixing parameter) per hour.

Applying the filtering MCMC to a reduced data representation using the 153 unique sighting histories, the complete model likelihood is calculated according to~(\ref{eqn:reducedFilterMCMC}), using~(\ref{eqn:filterMCMCmatrix}) to calculate the likelihood of each unique history.  Computation time is reduced to 20 minutes, which agrees with the expected speedup factor of $\frac{11,200}{153} \approx 73.2$.  Mixing also improves to produce a minimum sampling efficiency of 0.20 ESPS, a 70-fold improvement relative to the latent state MCMC.  This translates to 720 effective samples per hour, which may be considered practical.

NIMBLE's automated blocking procedure converges on seven blocks of parameters, ranging between two and five parameters each.  These seven blocks include 20 of the 21 parameters, leaving only one parameter for univariate sampling.  It is realistically unlikely that a practitioner would discover this blocking scheme through expert opinion or trial and error.  Runtime is comparable using this approach, but the joint sampling of correlated parameters gives a dramatic improvement in MCMC mixing.  The minimum sampling efficiency improves by an additional order of magnitude to 2.4 ESPS, or generating over 8,600 effective samples per hour.  This represents nearly a 1000-fold improvement over the latent state MCMC.

%



\section{Discussion}

We have studied alternate computational approaches for MCMC sampling of hierarchical models which include embedded discrete HMMs.  Traditional MCMC analysis of such models involves sampling the unknown (nuisance) latent states, whereas we propose filtering over latent states to calculate model likelihoods and limiting MCMC sampling to top-level parameters.  This introduces a computational trade-off: simplified MCMC sampling with the additional expense of filtering.  Through examples, we observe that worthwhile gains in sampling efficiency result from this approach.

Furthermore, the filtering MCMC permits a reduced representation of datasets with repeated observations.  This simplification is not possible when using traditional latent state MCMC, since each (possibly duplicated) observational history requires its own sequence of latent states.  When appropriate, combining our filtering MCMC with this reduced data representation provides an additional echelon of improvement in MCMC sampling efficiency, the extent of which is limited only by the degree of repetition in the initial data.

We note that the filtering MCMC approach forgoes generating posterior samples for latent states.  In some analyses the distribution of latent variables at a particular observational periods may be of interest, or otherwise may be used (for example) to estimate longevity distributions.  The inclusion of latent variables would also be necessary when used as explanatory variables in other parts of a hierarchical model \citep[\emph{e.g.},][]{risk2011robust}, or in the case of individual-specific covariates.  Our suggested approaches would not be appropriate in these analysis scenarios.

The analyses presented herein are facilitated by the NIMBLE package for R.  NIMBLE allows user-defined distribution functions to be used directly in hierarchical model specifications.  We define a multivariate distribution function parametrized by state transition and observation matrices, where the probability density evaluation routine implements discrete filtering to calculate likelihood values.  Models are specified using this distribution, which effectively embeds filtering into the model for the purposes of likelihood calculation.  NIMBLE's MCMC engine may then be applied to the resulting model to achieve the filtering MCMC.  We make use of NIMBLE's default MCMC as well as that resulting from automated parameter blocking.  The distinction of allowing programmable models and statistical algorithms, as compared to other statistical software, makes such analyses possible in NIMBLE.

\section*{Acknowledgements}
This work was supported by the NSF under grant DBI-1147230 and by support to DT from the Berkeley Institute for Data Science.  We thank Marc K\'{e}ry, Byron Morgan, and Michael Schaub for reviewing earlier versions of the manuscript.

\newpage
\printbibliography

\end{document}